\documentclass[a4paper]{jpconf}
\usepackage{amsmath}
\usepackage{graphicx}

\begin{document}

\title{Compact binary waveform recovery from the cross-correlated data of
two detectors by matched filtering with spinning templates}
\author{L. Ver\'{e}b$^{1,2}$, Z. Keresztes$^{1,2\star }$, P. Raffai$^{3\dag
} $, Zs. Udvari$^{1}$, M. T\'{a}pai$^{1,2}$, \ \ \  L. \'{A}. Gergely$^{1,2\ddag }$}

\begin{abstract}
We investigate whether the recovery chances of highly spinning waveforms by
matched filtering with randomly chosen spinning waveforms generated with the
LAL package are improved by a cross-correlation of the simulated output of
the L1 and H1 LIGO detectors. We find that a properly defined correlated
overlap improves the mass estimates and enhaces the recovery of spin angles.
\end{abstract}

\address{$^{1}$ Department of Theoretical Physics, University of Szeged, Tisza Lajos
krt 84-86, Szeged 6720, Hungary\\
$^{2}$ Department of Experimental Physics, University of Szeged, D\'{o}m t\'{e}r 9, Szeged 6720, Hungary\\
$^{3}$ Institute of Physics, Lor\'and E\"otv\"os University, P\'azmany P\'eter s 1/A, Budapest 1117, Hungary}

\ead{$^{\ast}$zkeresztes@titan.physx.u-szeged.hu\quad $^{\dag}$praffai@bolyai.elte.hu
\quad $^{\ddag }$gergely@physx.u-szeged.hu}

\section{Method}

Recovering a gravitational wave pattern in the noisy detector output is a
difficult problem. In this work we present our attempt to recover a
gravitational spinning waveform $h_{inj}$ immersed in S5 LIGO-like noise by
a modified matched filtering method. We calculate the overlap \cite{Overlap}
between the noisy injection $h_{n,i}$ and spinning templates $h_{template}$
as%
\begin{equation}
O\left[ h_{n,i},h_{template}\right] =\frac{\left\langle
h_{n,i}|h_{template}\right\rangle }{\sqrt{\left\langle
h_{n,i}|h_{n,i}\right\rangle \left\langle
h_{template}|h_{template}\right\rangle }}~,  \label{O}
\end{equation}%
with%
\begin{equation}
\left\langle h_{1}|h_{2}\right\rangle =4\mathit{Re}\int_{f_{\min }}^{f_{\max
}}\frac{\widetilde{h}_{1}\left( f\right) \widetilde{h}_{2}^{\ast }\left(
f\right) }{S_{n}\left( f\right) }df~,
\end{equation}%
where tilde and star denote Fourier transform and complex conjugate,
respectively, and $S_{n}\left( f\right) $ is the power spectral density of
the noise. We choose $f_{\min }=50$Hz and $f_{\max }=600$Hz, lying in the
best sensitivity band of the LIGO detectors, also the post-Newtonian
prediction for the waveform is quite accurate there. 
\begin{figure}[th]
\begin{center}
\includegraphics[height=7.0cm, angle=270]{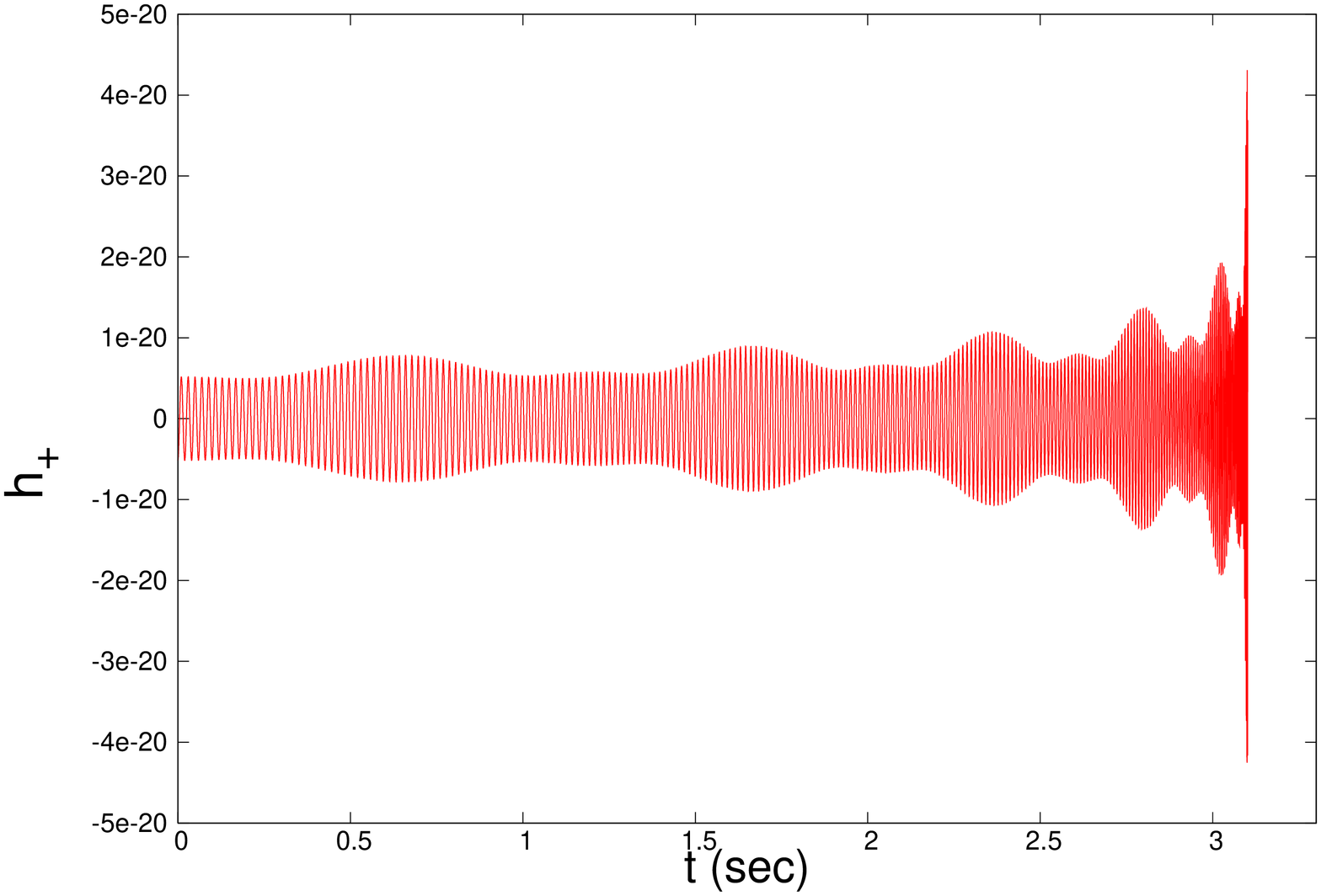} \hskip 0.5cm %
\includegraphics[height=7.0cm, angle=270]{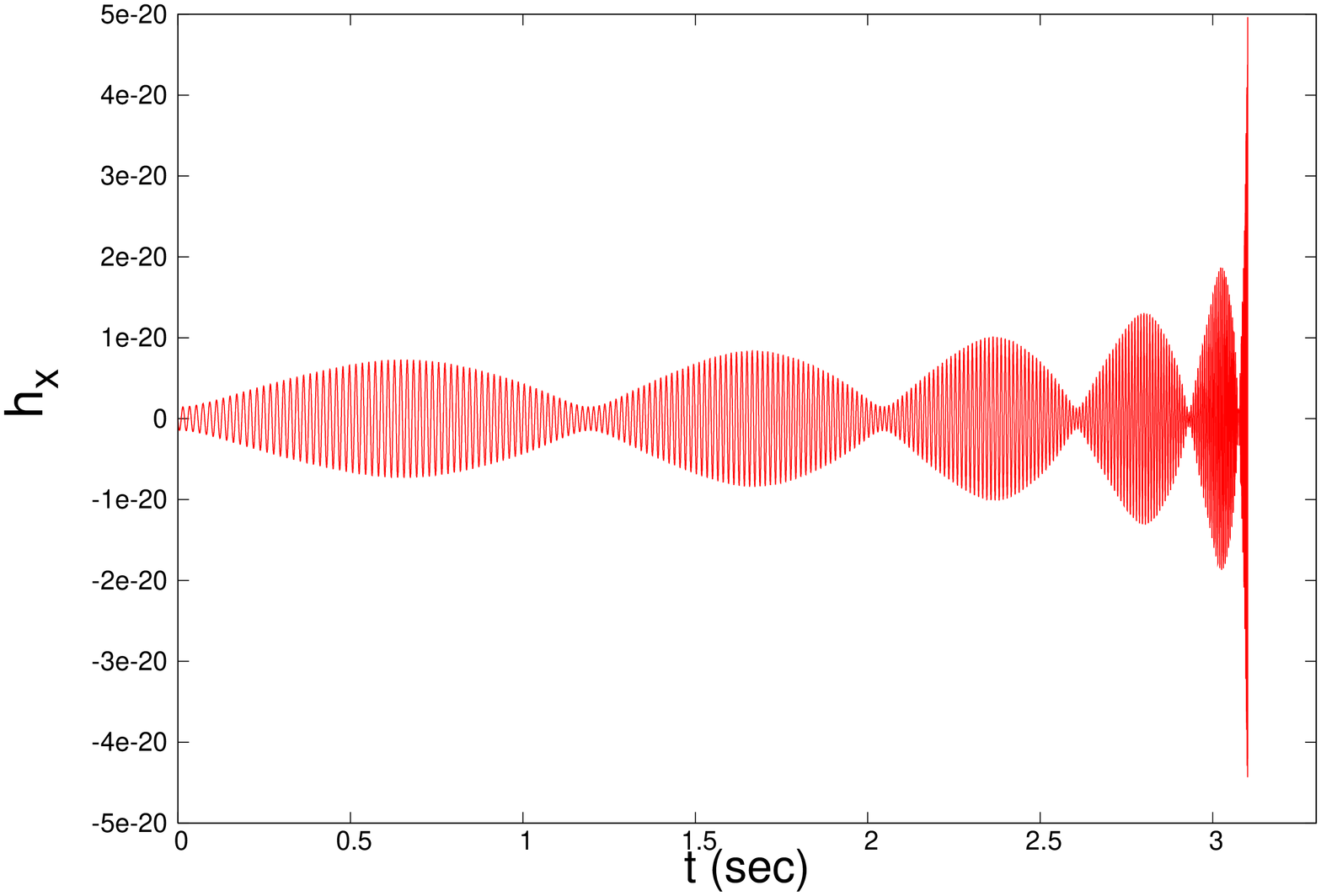}
\end{center}
\caption{ The polarization $h_{+}$ (left) and $h_{\times }$ (right) of the
spinning waveforms.}
\label{fig1}
\end{figure}

On theoretical grounds, infinite long data series would be required for
exact determination of the power spectrum. In order to achieve stability of
the power spectral density of order of 1\%, we would like to have at least
100 periods of the lowest frequency, therefore a minimal length of the
templates of 2 sec was imposed. We compute the above overlap both for the
Hanford and Livingston detectors ($O^{H}$ and $O^{L}$, respectively).

We also defined a \textit{correlated match}, the overlap $O_{corr}^{HL}$,
using the correlations of the Hanford and Livingston signals for $h_{n,i}$, $%
h_{template}$ and noises. (The correlation of $h_{1}$ and $h_{2}$ in Fourier
space is defined as $\widetilde{H}\left( f\right) =\widetilde{h}_{1}\left(
f\right) \widetilde{h}_{2}^{\ast }\left( f\right) $.) 
\begin{figure}[th]
\begin{center}
\includegraphics[height=7.0cm, angle=270]{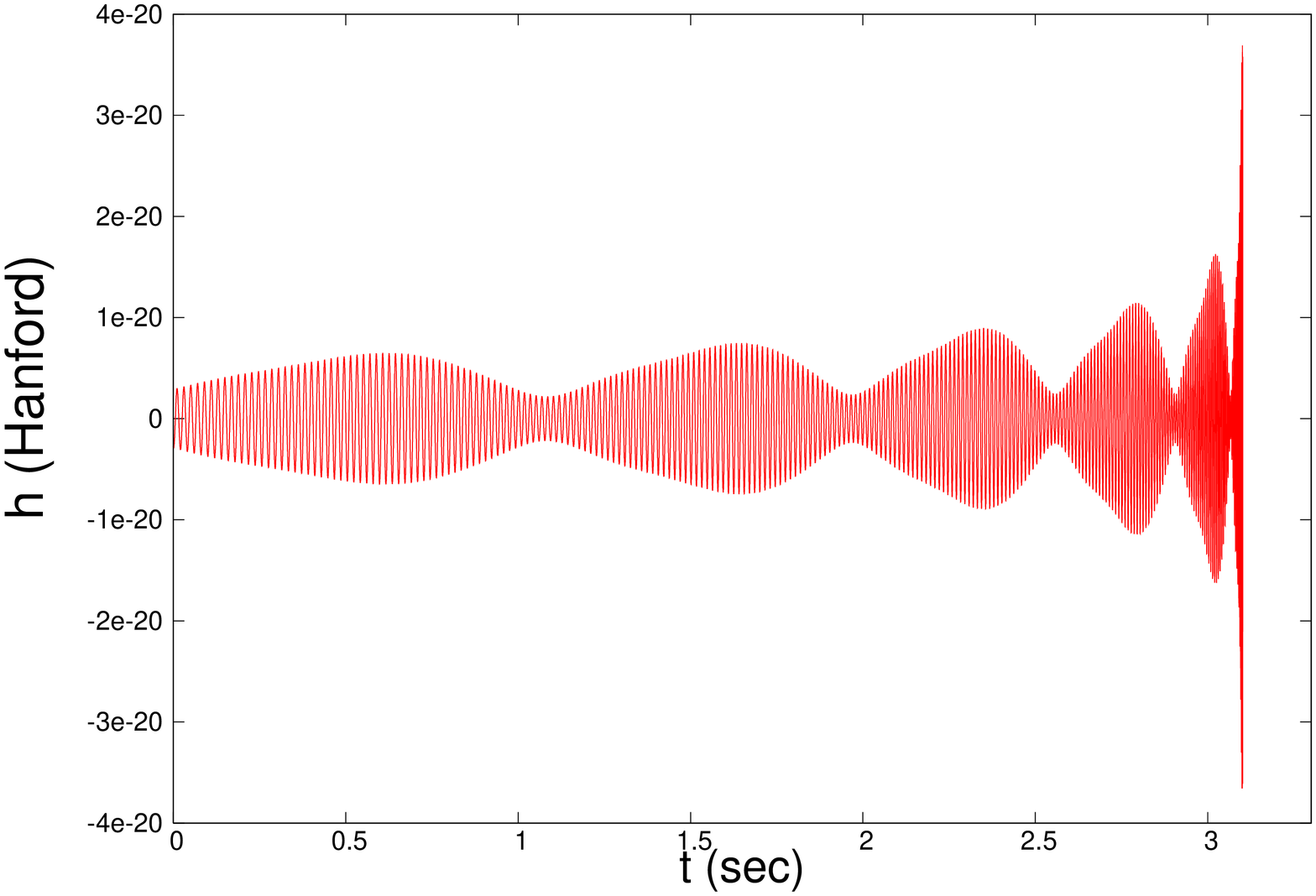} \hskip 0.5cm %
\includegraphics[height=7.0cm, angle=270]{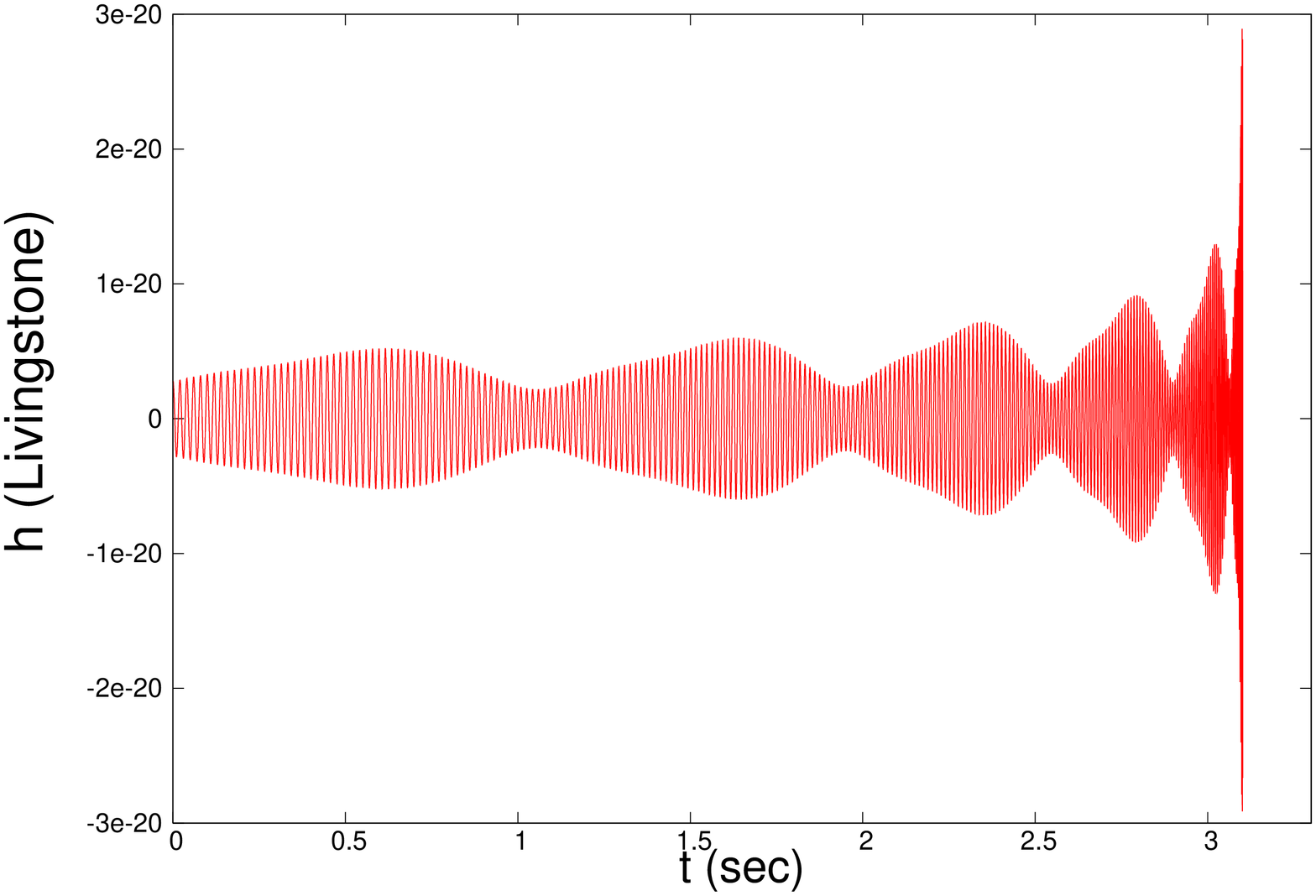} \vskip %
0.2cm \includegraphics[height=7.0cm, angle=270]{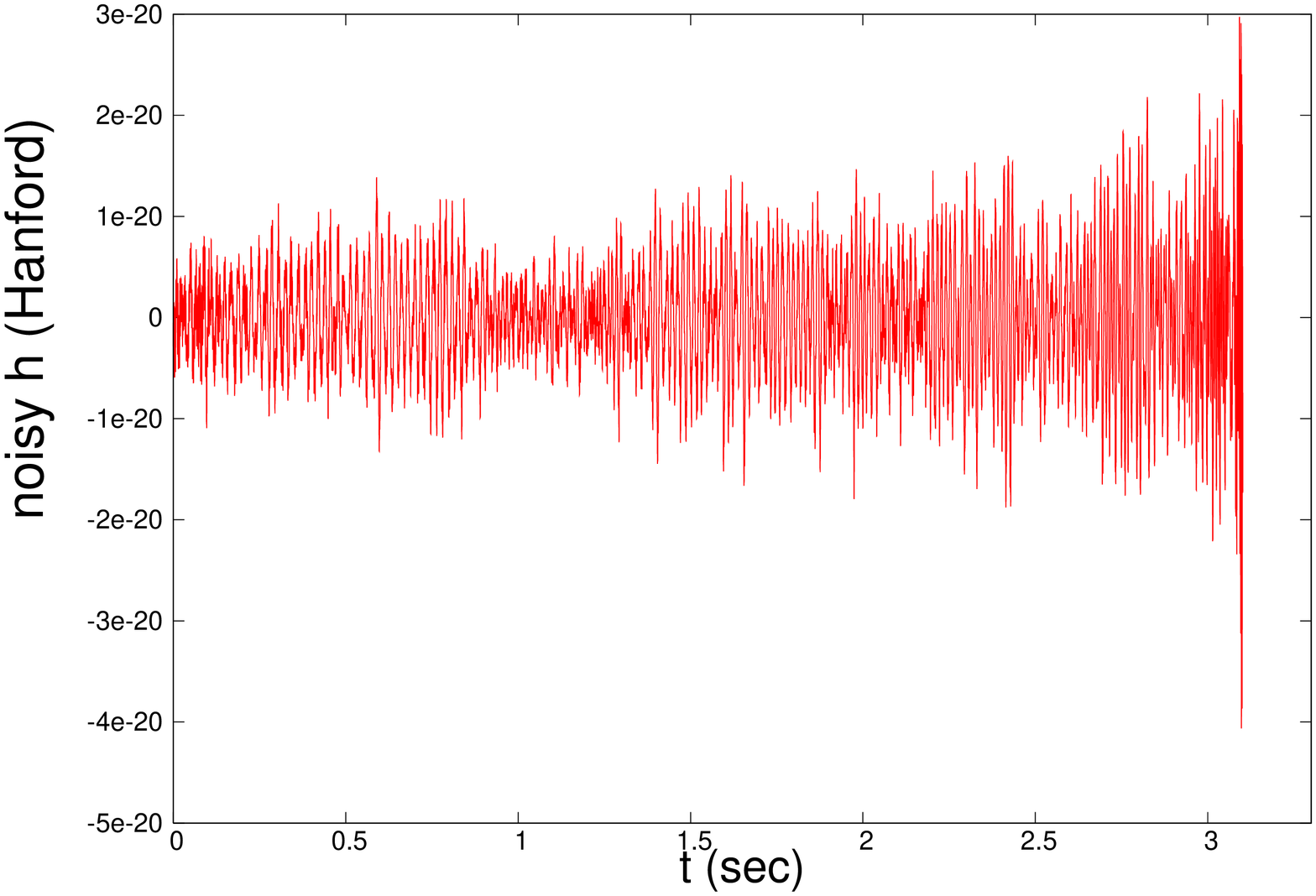} %
\hskip 0.5cm %
\includegraphics[height=7.0cm,angle=270]{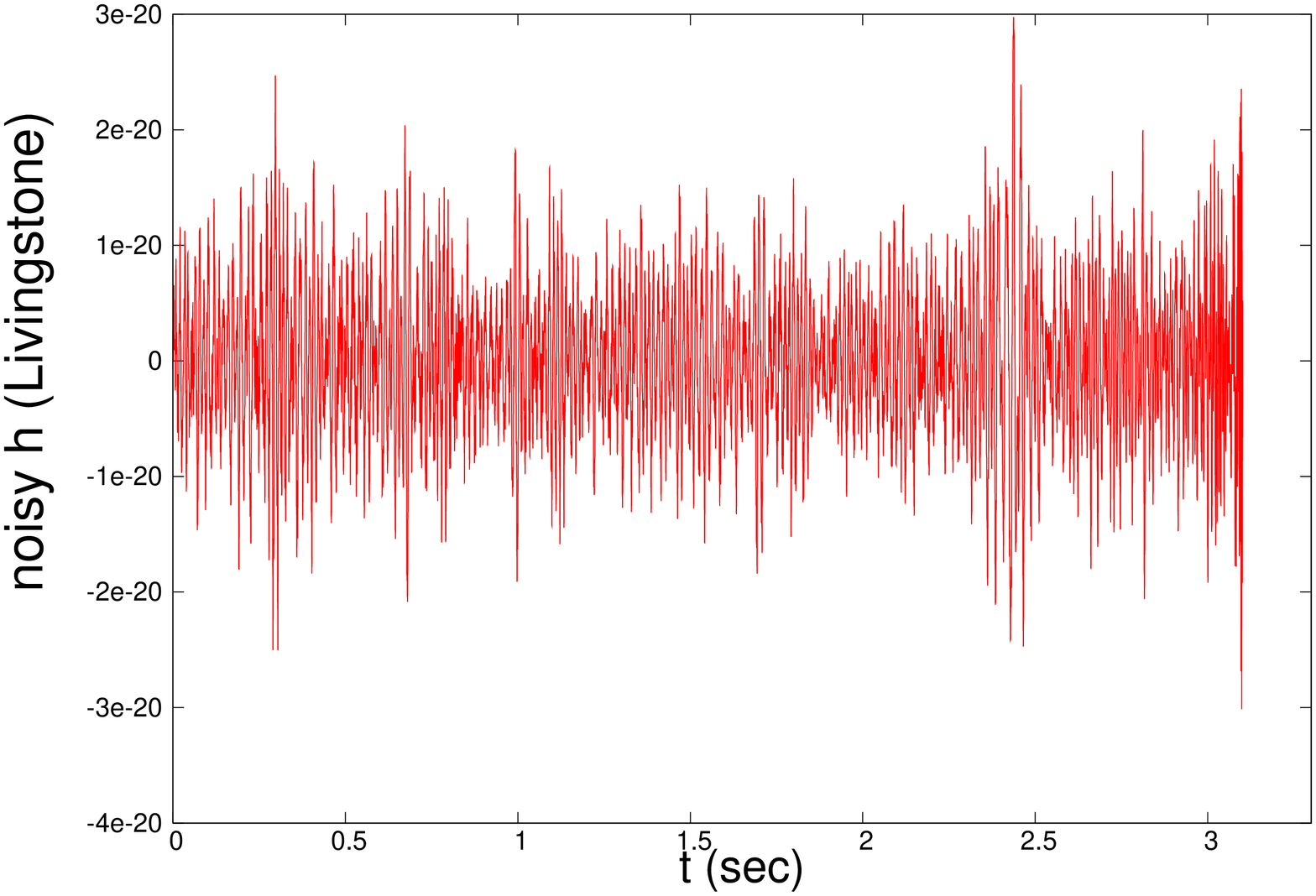}
\end{center}
\caption{ Spinning waveforms at Hanford (top left) and Livingston (top
right), mixed in LIGO S5-like noise (bottom).}
\label{fig2}
\end{figure}
\begin{figure}[th]
\begin{center}
\includegraphics[height=7.0cm, angle=270]{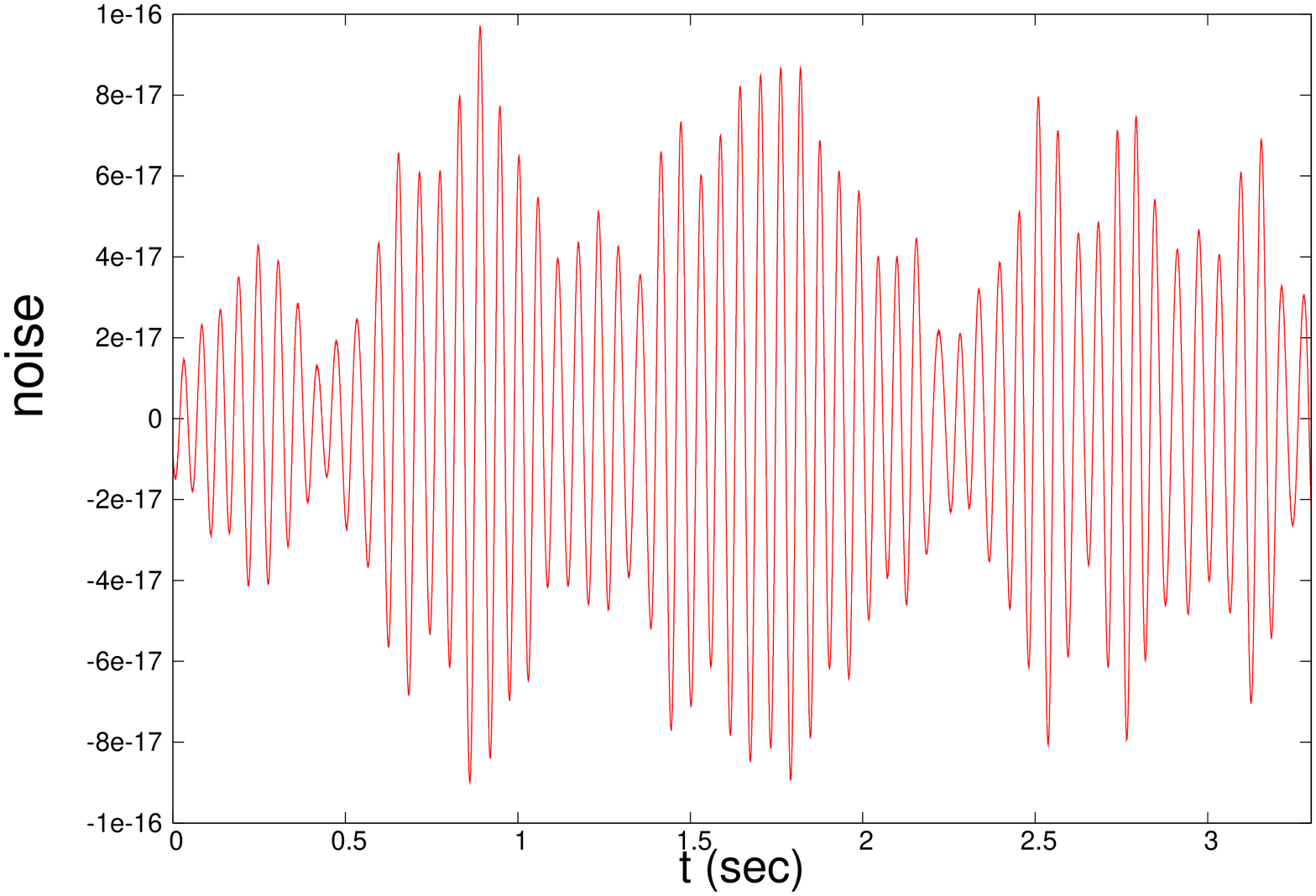} \hskip 0.5cm %
\includegraphics[height=7.0cm, angle=270]{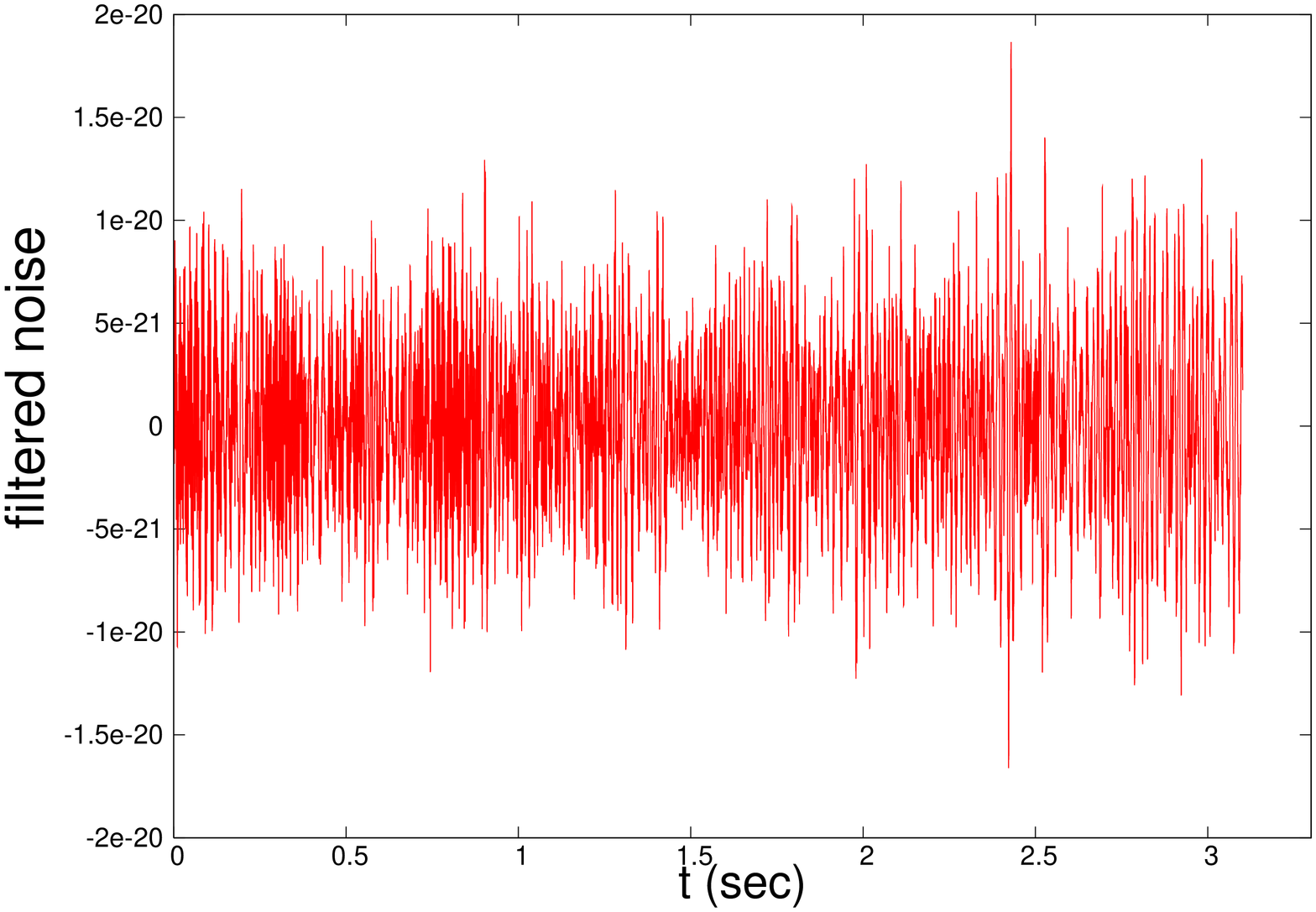}
\end{center}
\caption{ Characteristic unfiltered (left) and filtered (right) LIGO noises.}
\label{fig3}
\end{figure}

For the two detectors we also employed the quantity $O^{HL}$, according to
Ref. \cite{SpinSpiral}:%
\begin{equation}
O^{HL}=\frac{\left\langle h_{n,i}|h_{template}\right\rangle
_{H}+\left\langle h_{n,i}|h_{template}\right\rangle _{L}}{\sqrt{\left(
\left\langle h_{n,i}|h_{n,i}\right\rangle _{H}+\left\langle
h_{n,i}|h_{n,i}\right\rangle _{L}\right) \left( \left\langle
h_{template}|h_{template}\right\rangle _{H}+\left\langle
h_{template}|h_{template}\right\rangle _{L}\right) }}~.
\end{equation}%
Therefore we could compare four type of matches.

The spinning waveforms, based on Ref. \cite{BCV2}, were computed by using
SpinTaylor Ref. \cite{SpinTaylor}, while the antenna functions based on Ref. 
\cite{Antenna} with the XLALComputeDetAMResponse() function in DetResponse.c
under the LAL package Ref. \cite{SpinTaylor}.

\begin{table}[t]
\caption{The parameters of the gravitational wave polarizations ($%
h_{+},h_{\times }$) of the injected signal. Masses $m_{i}$, magnitude of the
dimensionless spins $\protect\chi _{i}$, initial direction of the spin
vectors given by $\cos \protect\kappa _{i}$ and $\protect\psi _{i}$ in the
frame with the line of sight on the $z$-axis, and the initial orbital
angular momentum $\mathbf{L}_{\mathbf{N}}$ in the $x$-$z$ plane, spanning
the angle $\Theta $, and the distance $d_{L}$ of the source. The initial
phase is $0$; the initial time being also fixed to $0$.}$%
\begin{tabular}{c|c|c|c|c|c|c|c|c|c|c}
name & $m_{1}(M_{\odot })$ & $m_{2}(M_{\odot })$ & $\chi _{1}$ & $\chi _{2}$
& $\cos \kappa _{1}$ & $\psi _{1}$ & $\cos \kappa _{2}$ & $\psi _{2}$ & $%
\Theta $ & $d_{L}\left( Mpc\right) $ \\ \hline
injection & $3.553$ & $3.358$ & $0.983$ & $0.902$ & $0.984$ & $1.109$ & $%
0.978$ & $0.957$ & $1.430$ & $1$%
\end{tabular}%
$%
\label{table1}
\end{table}
\begin{table}[t]
\caption{The angles $\protect\theta $, $\protect\varphi $ and $\protect\psi $
(polarization angle) give the relation between a fictitious Earth-centered
detector and the source frame. For the actual detector positions further
rotation have to be taken into account \protect\cite{Antenna}. These angles
are also necessary for computing the antenna functions ($F_{+},F_{\times }$%
). The gravitational signal is $h=h_{+}F_{+}+h_{\times }F_{\times }$.}
\begin{center}
$%
\begin{tabular}{c|c|c|c}
name & $\varphi $ & $\theta $ & \multicolumn{1}{|c}{$\psi $} \\ \hline
injection & $3.657$ & $0.278$ & \multicolumn{1}{|c}{$0.000$}%
\end{tabular}%
$%
\end{center}
\label{table2}
\end{table}

Table \ref{table1} contains the parameters of the injected waveform. On Fig %
\ref{fig1} the two polarizations of this waveform $h_{+}$ and $h_{\times }$
are shown. The parameters characterizing the source and the detector
orientation, which are necessary for computing the antenna functions $F_{+}$
and $F_{\times }$ are given in Table \ref{table2}. The waveforms appearing
at the Hanford and Livingston detectors are plotted on Fig \ref{fig2} (top
line), while the bottom line on Fig \ref{fig2} shows the signals immersed in
S5 LIGO-like noise. Some characteristic unfiltered and filtered noises can
be seen on Fig \ref{fig3}.

First we calculated the four type of overlaps between the noisy injection
and the injected signal itself, finding $O^{H}\left[ h_{n,i},h_{injection}%
\right] =0.8392$, $O^{L}\left[ h_{n,i},h_{injection}\right] =0.9064$, $O^{HL}%
\left[ h_{n,i},h_{injection}\right] =0.8662$ and $O_{corr}^{HL}\left[
h_{n,i},h_{injection}\right] =0.8646$. Next we defined the following
auxiliary quantity for all overlaps%
\begin{equation}
\sigma =\left\vert 1-\frac{O\left[ h_{n,i},h_{template}\right] }{O\left[
h_{n,i},h_{injection}\right] }\right\vert ~,
\end{equation}%
vanishing for the injected template. We searched for templates with the
various $\sigma $-s less then $0.1$. 
\begin{figure}[th]
\includegraphics[height=16cm, angle=270]{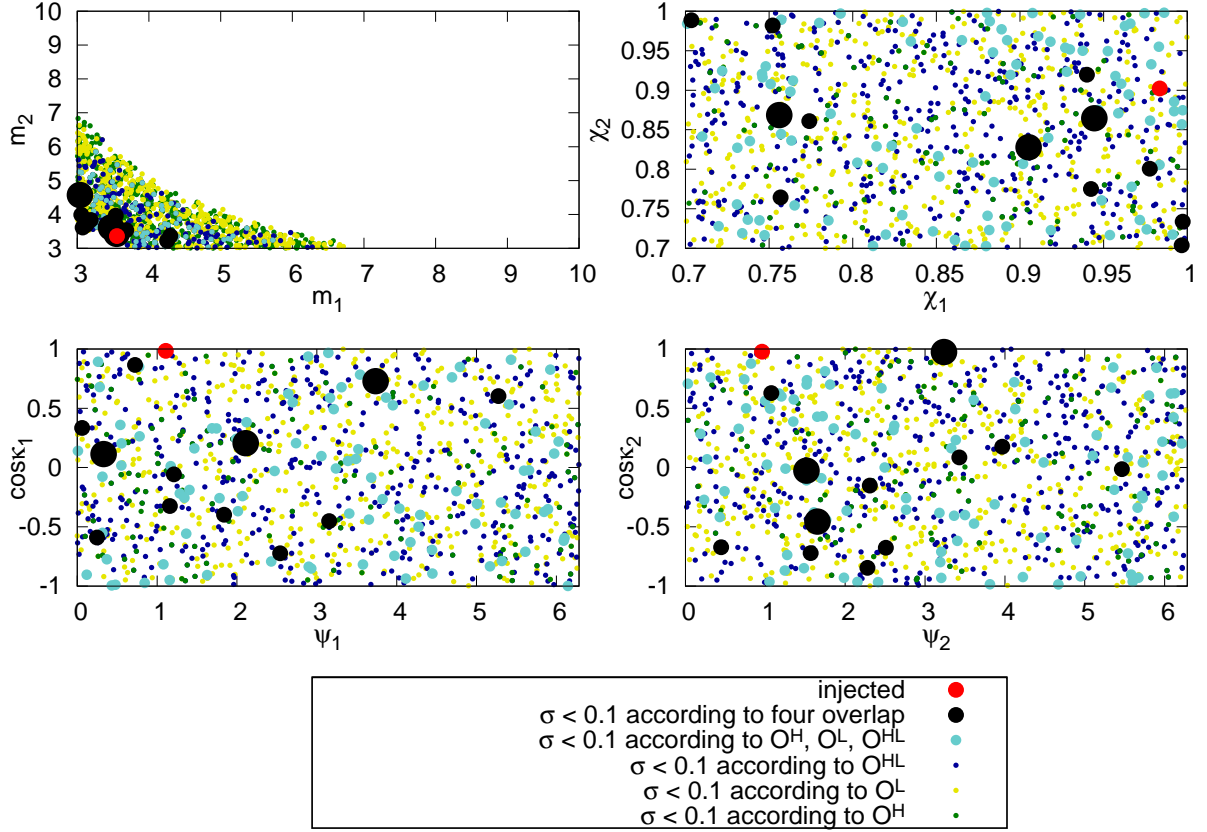}
\caption{The parameters of \ the injected signal (red) and the templates
having $\protect\sigma <1$ according to $O^{H}$ (green), $O^{L}$ (yellow), $%
O^{HL}$ (dark blue) and all of these (light blue) and when the correlated
overlap was also taken into account (black). The big dots among the black
represent those templates that has the smallest values accordint to the
correlated overlap.}
\label{fig4}
\end{figure}

The templates were chosen with parameters in the ranges: masses $m_{i}\in $ $%
3\div 10$ M$_{\odot }$; dimensionless spins $\chi _{i}\in 0.7\div 1$; spin
angles $\cos \kappa _{i}$ and $\psi _{i}$ random; distance $d_{L}$, angles $%
\Theta $, $\varphi $, $\theta $ and $\psi $ are fixed identically to the
values given in Tables \ref{table1} and \ref{table2} for the injection,
respectively. We also assumed as known the time of signal arrival to the
detector. Therefore we varied two mass and six spin parameters altogether. 
\begin{figure}[th]
\includegraphics[height=7.0cm, angle=270]{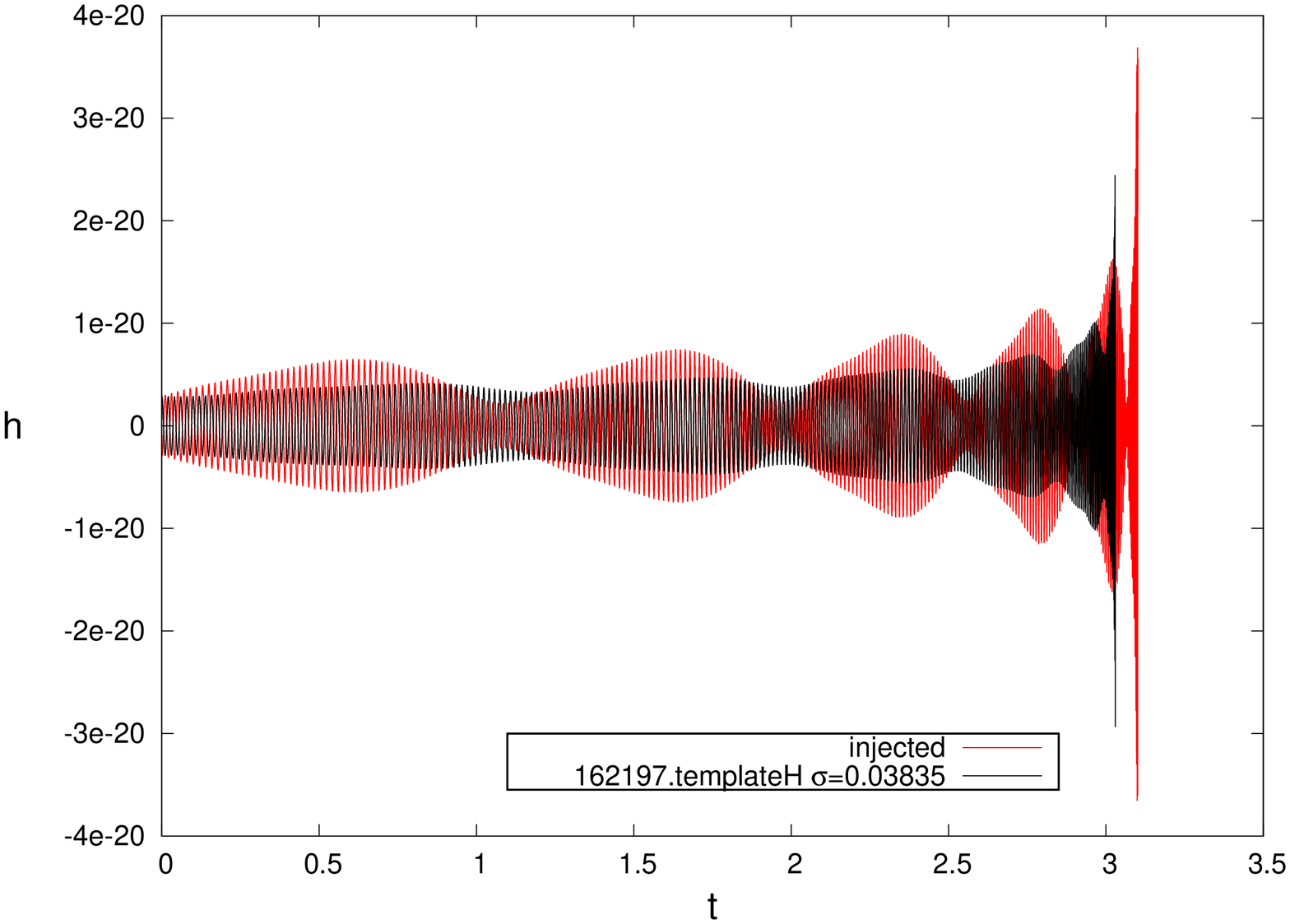} \hskip0.5cm %
\includegraphics[height=7.0cm, angle=270]{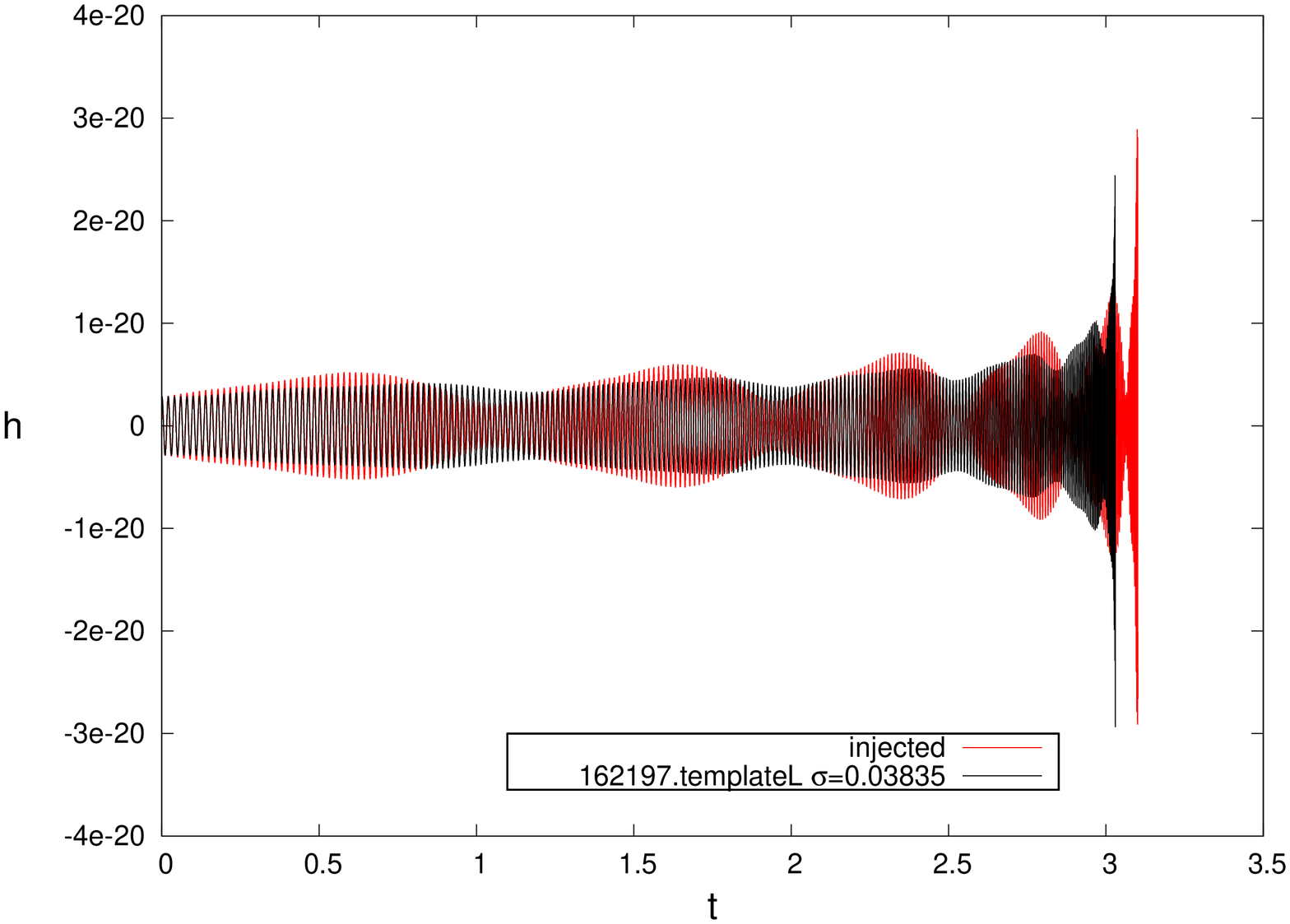} \vskip0.2cm %
\includegraphics[height=7.0cm, angle=270]{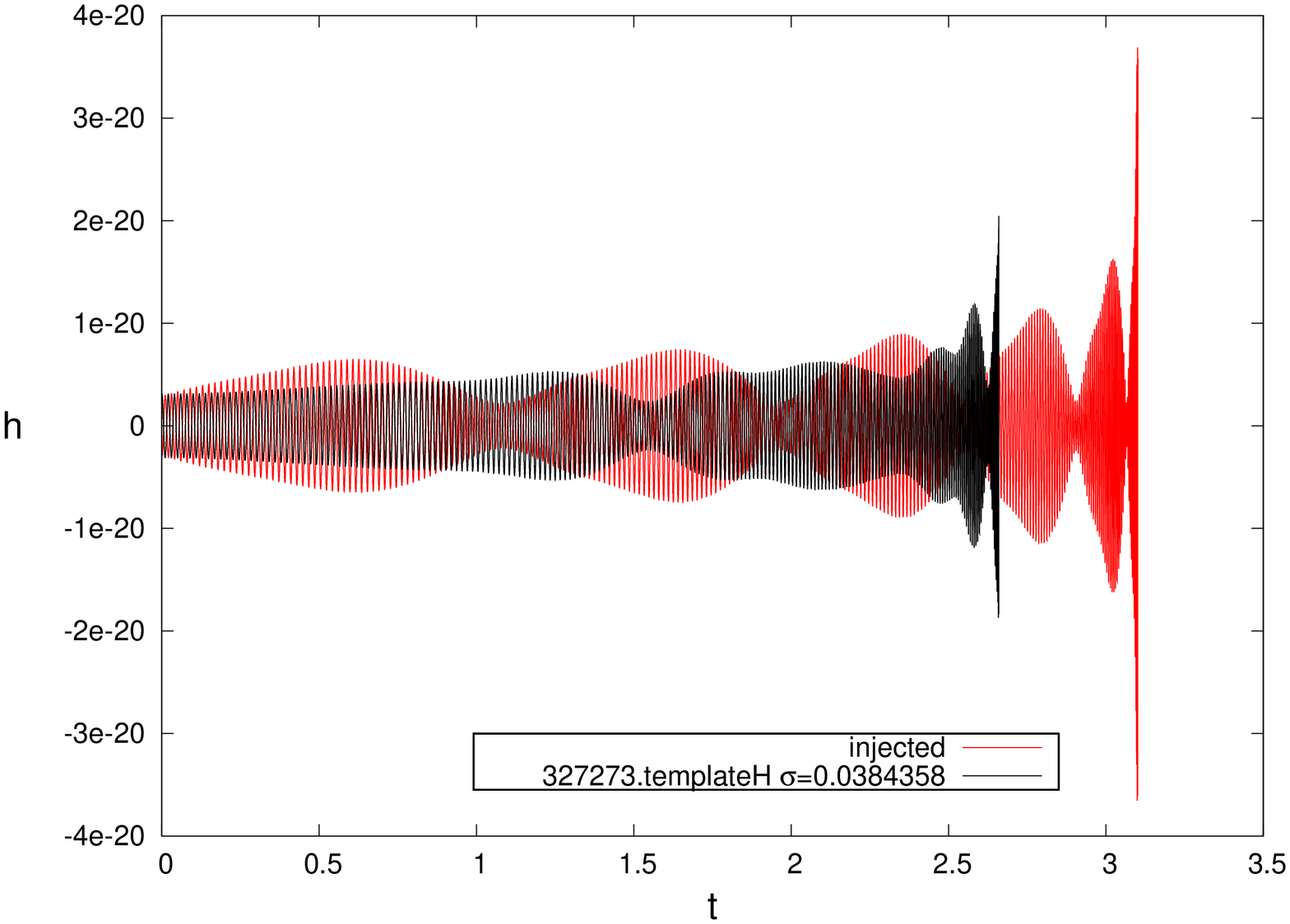} \hskip0.5cm %
\includegraphics[height=7.0cm,angle=270]{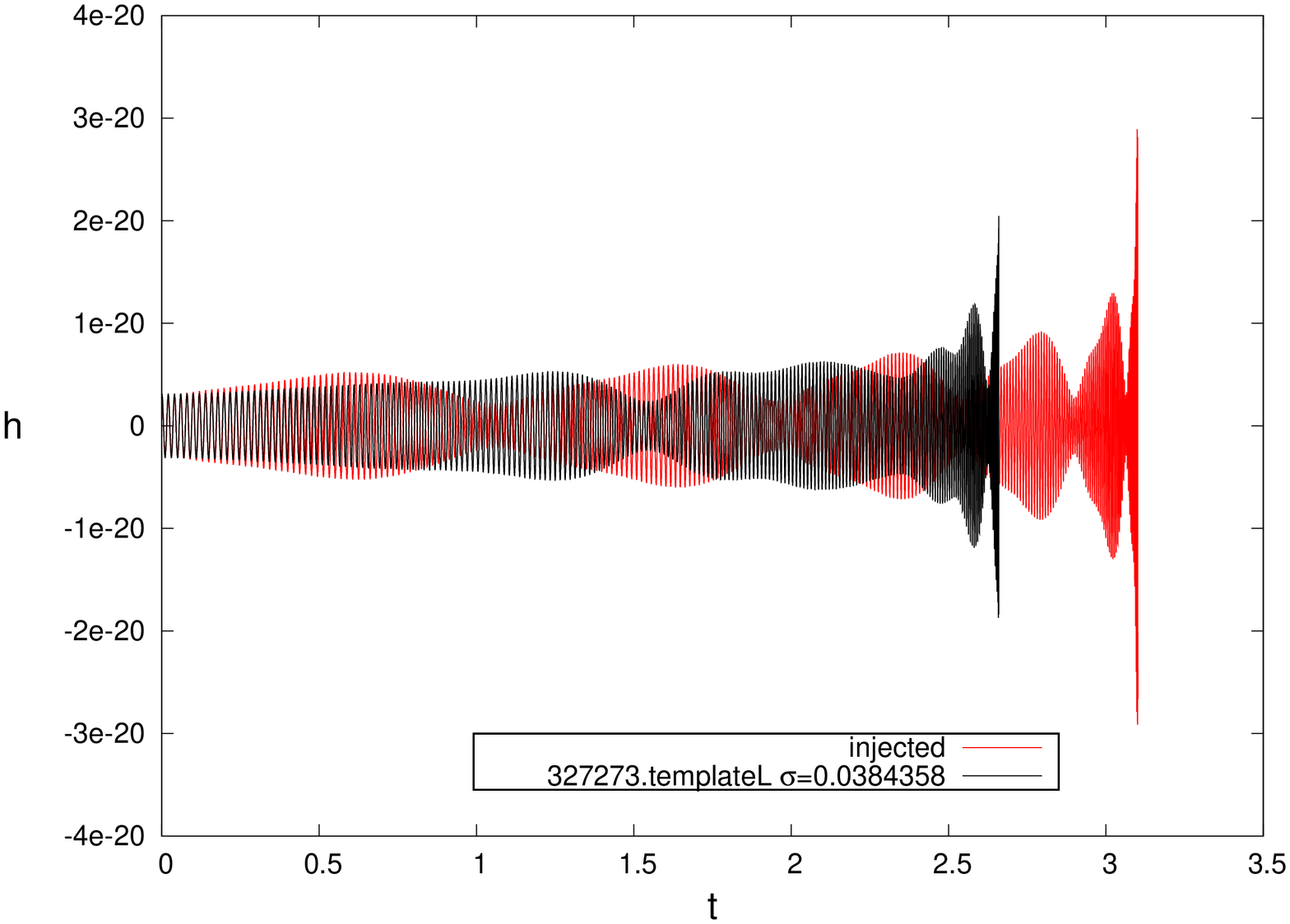} \vskip0.2cm %
\includegraphics[height=7.0cm, angle=270]{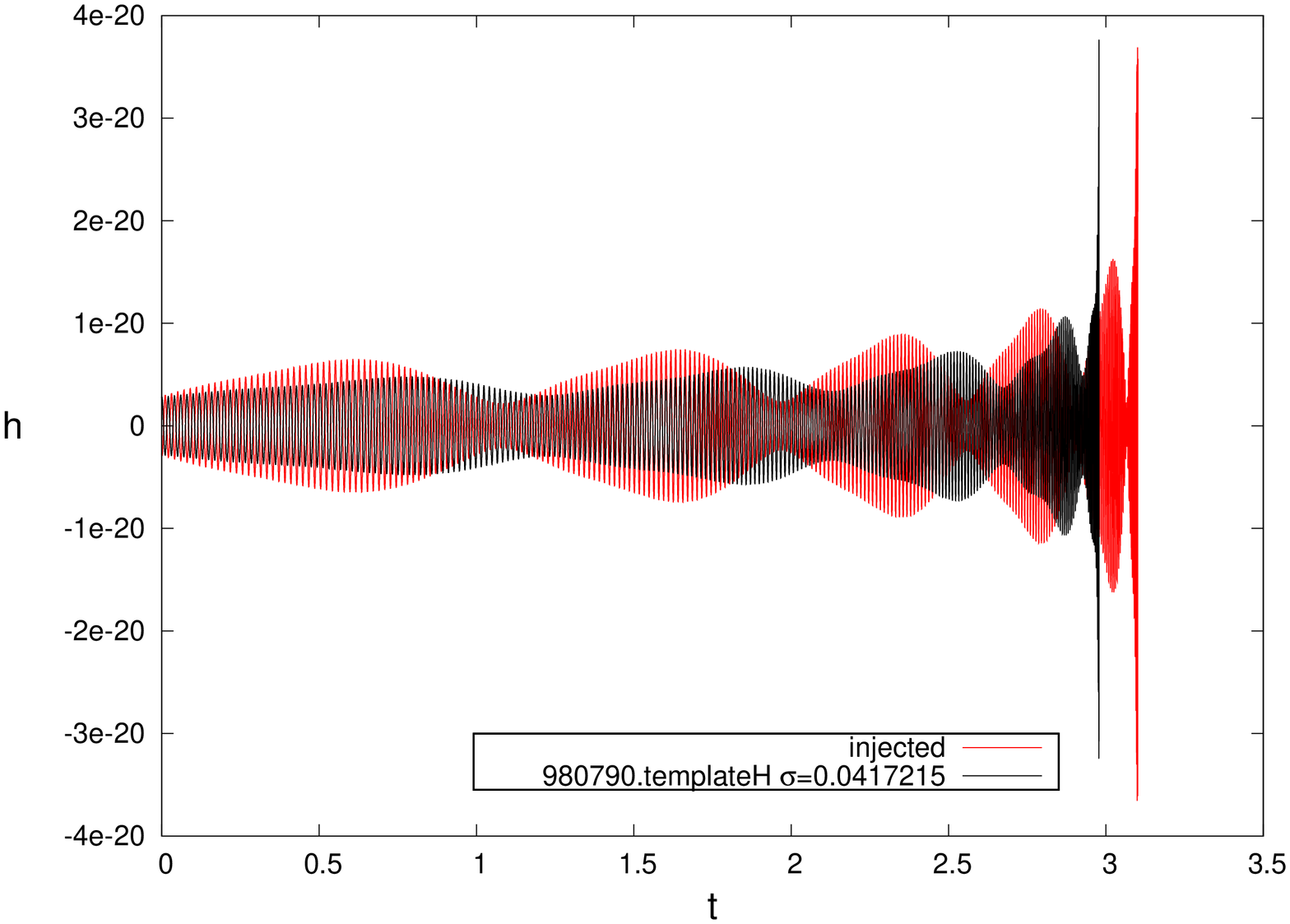} \hskip0.5cm %
\includegraphics[height=7.0cm, angle=270]{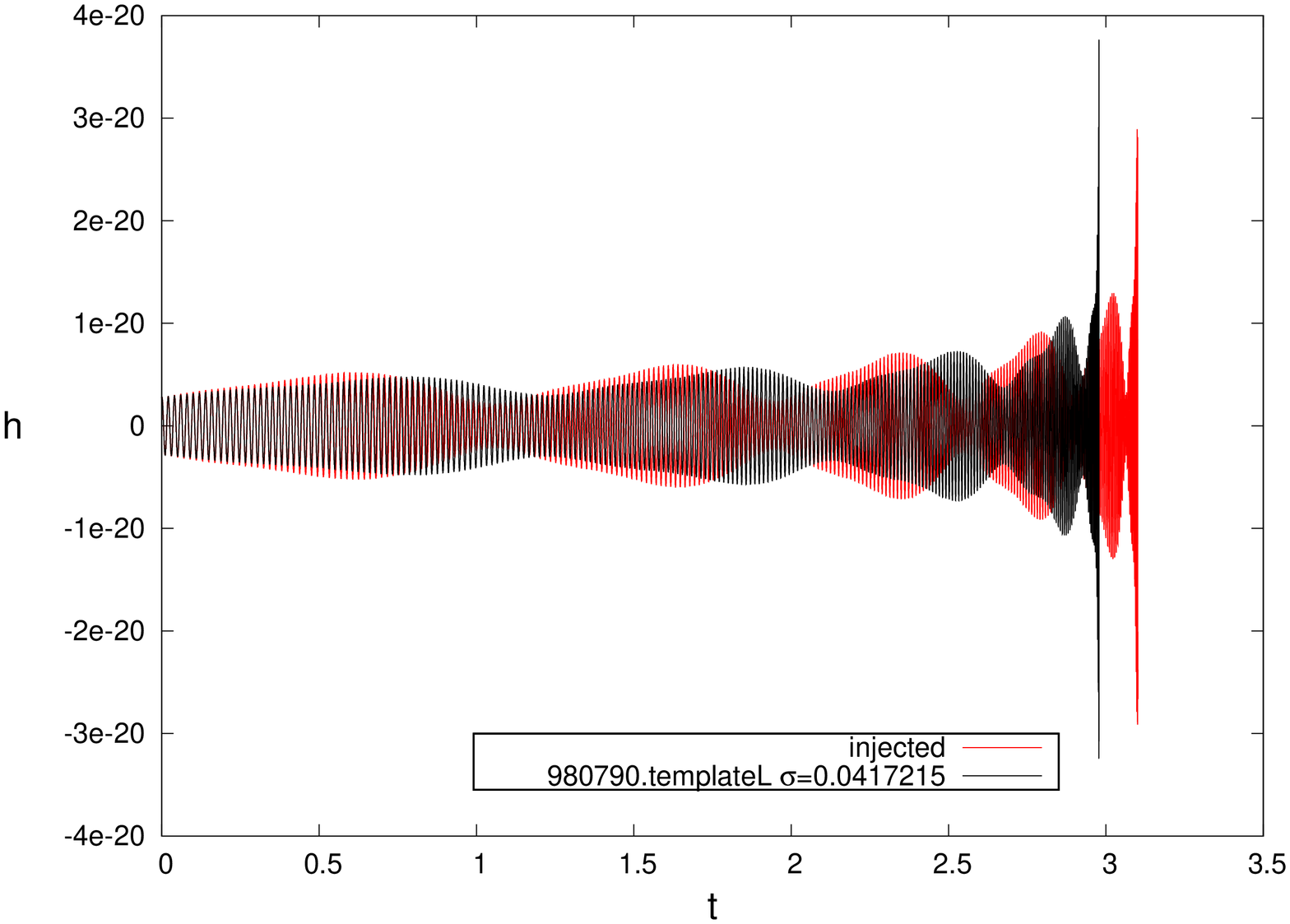} 
\caption{Best three templates (black) as they would appear at Hanford (left)
and Livingston (right), compared to the respective injected signals (red).}
\label{fig5}
\end{figure}

\section{Results}

Our analysis is based on matching more than one million templates. The
templates with any of the $\sigma ^{H}$, $\sigma ^{L}$, $\sigma ^{HL}$ $<0.1$
were selected and represented on Fig \ref{fig4} in the parameter planes ($%
m_{1},m_{2}$), ($\chi _{1},\chi _{2}$) and ($\cos \kappa _{i},\psi _{i}$)
for both spins. The green, yellow and navy dots (for grayscale see the
legend) represent templates with required values of $\sigma ^{H}$, $\sigma
^{L}$and $\sigma ^{HL}$, respectively; templates with all three values of $%
\sigma $ below the threshold are plotted with larger turqoise dots. Twelve
even larger black dots show templates with all four $\sigma $-s (including $%
\sigma _{corr}^{HL}$) below the threshold. The three largest of them have
the lowest value of $\sigma _{corr}^{HL}$. The~parameters of the best three
templates are shown in Table \ref{table3}, and they are plotted on Fig \ref%
{fig5} as they would appear at the Hanford and Livingston detectors.

\textit{Discussion}: The masses are reasonably well recovered, although
sligtly overestimated with any of the $\sigma ^{H}$, $\sigma ^{L}$, $\sigma
^{HL}$. The additional monitoring of the correlated match $\sigma
_{corr}^{HL}$ imposes however a selection effect which \textit{improves the
estimation of the masses }(top left panel of Fig \ref{fig4}). While the
recovery of the spin magnitudes is still problematic (top right panel of Fig %
\ref{fig4}), the estimation of the spin angles seems slightly improved by
the use of of the correlated match $\sigma _{corr}^{HL}$. (Black dots
exhibit a belt-like structure on both bottom panels of Fig \ref{fig4}.) How
relevant is this feature statistically is currently under investigation. 
\begin{table}[t]
\caption{The parameters of the templates represented by the four big black
points on Fig \protect\ref{fig4}.}$%
\begin{tabular}{c|c|c|c|c|c|c|c|c|}
name & $m_{1}(M_{\odot })$ & $m_{2}(M_{\odot })$ & $\chi _{1}$ & $\chi _{2}$
& $\cos \kappa _{1}$ & $\psi _{1}$ & $\cos \kappa _{2}$ & $\psi _{2}$ \\ 
\hline
162197.template & $3.471$ & $3.612$ & $0.905$ & $0.828$ & $0.114$ & $0.330$
& $-0.026$ & $1.515$ \\ 
327273.template & $3.038$ & $4.572$ & $0.944$ & $0.864$ & $0.729$ & $3.736$
& $-0.455$ & $1.652$ \\ 
980790.template & $3.549$ & $3.426$ & $0.756$ & $0.868$ & $0.206$ & $2.110$
& $0.974$ & $3.237$ \\ 
281270.template & $3.057$ & $3.989$ & $0.703$ & $0.989$ & $-0.057$ & $1.209$
& $-0.721$ & $1.569$%
\end{tabular}%
\ $%
\label{table3}
\end{table}

\textit{Acknowledgements}: This work was supported by the Pol\'{a}nyi
Program of the Hungarian National Office for Research and Technology (NKTH)
and the Hungarian Scientific Research Fund (OTKA) grant no. 69036.

\end{document}